\documentclass[
onecolumn,
 amsmath,amssymb,
 aps,
 nofootinbib,
]{revtex4-2}

\usepackage{graphicx}
\usepackage{bm}

\usepackage{xcolor}

\begin{document}

\title{Friedmann equations and cosmic bounce in a modified cosmological scenario}

\author{Ana Alonso-Serrano}
\email{ana.alonso.serrano@aei.mpg.de}
\affiliation{Max-Planck-Institut f\"ur Gravitationsphysik (Albert-Einstein-Institut), \\Am M\"{u}hlenberg 1, 14476 Potsdam, Germany}

\author{Marek Li\v{s}ka,}
\email{liska.mk@seznam.cz}
\affiliation{Institute of Theoretical Physics, Faculty of Mathematics and Physics, Charles University,
	V Hole\v{s}ovi\v{c}k\'{a}ch 2, 180 00 Prague 8, Czech Republic}
\affiliation{Max-Planck-Institut  f\"ur  Gravitationsphysik  (Albert-Einstein-Institut), \\
	Am M\"{u}hlenberg 1, 14476 Potsdam, Germany}

\author{Antonio Vicente-Becceril}
\email{antonio.vicente-becerril@polytechnique.edu}
\affiliation{Ecole Polytechnique,\\
Route de Saclay, F-91128 Palaiseau Cedex, France}

\begin{abstract}
In this work we present a derivation of modified Raychaudhuri and Friedmann equations from a phenomenological model of quantum gravity based on the thermodynamics of spacetime. Starting from general gravitational equations of motion which encode low-energy quantum gravity effects, we found its particular solution for homogenous and isotropic universes with standard matter content, obtaining a modified Raychaudhuri equation. Then, we imposed local energy conservation and used a perturbative treatment to derive a modified Friedmann equation. The modified evolution in the early universe we obtained suggests a replacement of the Big Bang singularity by a regular bounce. Lastly, we also briefly discuss the range of validity of the perturbative approach and its results.
\end{abstract}

\maketitle


\section{Introduction}

As of today, a complete theory of quantum gravity remains elusive. Nevertheless, various approaches proposing a quantisation of gravity have been applied in highly symmetric settings, which allow to obtain definite results. The simplest such setting is a homogeneous and isotropic cosmological spacetime, whose dynamics is fully described by a single function of the cosmological time $t$, the scale factor $a\left(t\right)$. The importance of applying quantum gravity to cosmology lies, on the one side, in finding potentially testable predictions of quantum gravitational effects in the early universe. On the other side, the lessons learned in this relatively simple case can provide clues to the general behaviour of quantum gravity.
		
A well explored approach of this kind is loop quantum cosmology~\cite{Bojowald:2001,Ashtekar:2006,Martin-Benito:2009}. It employs quantisation techniques inspired by loop quantum gravity in a symmetry reduced setting (minisuperspace) corresponding, in its simplest case, to the homogeneous, isotropic cosmological spacetime. The most notable prediction of loop quantum cosmology is the replacement of the Big Bang singularity by a regular bounce, i.e. a point at which the universe stops contracting and starts to expand \footnote{According to improved dynamics of loop quantum cosmology, the bounce occurs when the density reaches the order of the Planck scale, $\rho_{\text{crit}}\approx0.41\rho_{\text{P}}$~\cite{Agullo:2016}. The value of $\rho_{\text{crit}}$ depends on the Barbero-Immirzi parameter and in this case, it is fixed by demanding that one correctly reproduces Bekenstein expression for black hole entropy in loop quantum gravity.}. This approach has some drawbacks, mainly concerned with its relation to the full theory, values of the constants parametrising the theory, and also its regime of validity. Nevertheless, it presents solid predictions for the early universe cosmology (For a critical analysis of the current state of loop quantum cosmology see, e.g.~\cite{Bojowald:2020}).

Here, we propose a novel approach to quantum gravitational effects in cosmology. It employs general low energy quantum gravitational dynamics we previously derived from thermodynamics of spacetime~\cite{Alonso:2020}. The key assumption of our approach is that local, observer-dependent causal horizons posses Bekenstein entropy of the same form as black hole horizons. Then, thermodynamic equilibrium conditions for local causal horizons encode the equations governing gravitational dynamics~\cite{Jacobson:1995} \footnote{Let us remark that the equilibrium conditions can also be phrased  fully in terms of quantum entanglement~\cite{Jacobson:2016}. This allows a more rigorous derivation of equations of gravitational dynamics from equilibrium conditions, details of which are not relevant for our current purposes (for an extensive discussion, see~\cite{Alonso:2020}). The leading order corrections to entanglement entropy can be rigorously shown to be logarithmic, as required by our approach~\cite{Mann:1998,Solodukhin:2011}.}. In the mentioned approach, to study low energy quantum gravitational dynamics, we consider the leading order correction to Bekenstein entropy, i.e., a term logarithmic in horizon area $\mathcal{A}$
\begin{equation}\label{modi-entrop}
S_{\text{B}}=\frac{k_{\text{B}}\mathcal{A}}{4l_{\text{P}}^2}+18\pi Dk_{\text{B}}\ln{\left(\frac{\mathcal{A}}{\mathcal{A}_0}\right)}+ O\left(\frac{k_{\text{B}}l_{\text{P}}^2}{\mathcal{A}}\right),
\end{equation}
where $k_{\text{B}}$ is the Boltzmann constant, $l_{\text{P}}$ is the Planck length, $\mathcal{A}_0$ is an arbitrary constant with dimensions of area, and $D\in \mathbb{R}$ is a dimensionless parameter whose value and sign depend on the particular model in which we calculate the modified Bekenstein entropy (since $D$ appears in the modified gravitational equations, we write the proportionality constant in entropy here as $18\pi D$). This form of Bekenstein entropy is nearly universally predicted by the diverse candidate theories to quantum gravity (e.g. loop quantum gravity~\cite{Kaul:2000}, string theory~\cite{Sen:2013}, AdS/CFT correspondence~\cite{Faulkner:2013}), as well as by various model independent phenomenological arguments~\cite{Adler:2001,Gour:2003,Davidson:2019} and by entanglement entropy calculations~\cite{Mann:1998}.	

The modified equations for gravitational dynamics were derived starting from the modified Bekenstein entropy~\eqref{modi-entrop} and using equilibrium conditions applied to local causal diamonds~\cite{Alonso:2020}. These equations incorporate in that way low energy quantum gravitational corrections fully determined by the value of the proportionality constant $D$. Given the universality of equation~\eqref{modi-entrop}, the resulting equations do not depend on any specific assumptions about the fundamental theory of quantum gravity. Notably, for a given theory of quantum gravity, $D$ tends to be a universal constant independent of any arbitrary parameters~\cite{Solodukhin:2011}. For instance, $D$ is a universal constant in loop quantum gravity, whereas the proportionality constant in the leading order term in black hole entropy depends on the (in principle arbitrary) Barbero-Immirzi parameter~\cite{Kaul:2000}. Likewise, the logarithmic correction appearing in entanglement entropy calculations does not depend on the regularisation scheme employed~\cite{Solodukhin:2011} (unlike the leading term). Therefore, our approach to low energy quantum gravitational dynamics is not only model independent, but should avoid even the dependence on parameters of the full quantum gravity (unlike loop quantum cosmology, whose predictions depend on the Barbero-Immirzi parameter and the area gap).

In this work, we specialise these general modified equations of motion to the case of homogeneous, isotropic cosmological spacetimes. In the present analysis, we approximate matter content of the universe as a classical perfect fluid (in general multi-component). We find that for $D>0$ our model perturbatively yields predictions equivalent with the effective dynamics of loop quantum cosmology~\cite{Ashtekar:2015} and, therefore, suggests the replacement of the Big Bang singularity with a regular bounce. Due to the considered perturbative expansion, the domain of validity of our equations does not extend all the way to the bounce. Nevertheless, we find that the dynamics continue to behave consistently with loop quantum cosmology picture almost all the way to the Planck scale. Given the model independence of our approach and the thermodynamic motivation for it, our results suggests that the loop quantum cosmology picture of the early universe (up to the Planck scale) may be qualitatively accurate independently of the final theory of quantum gravity.

The paper is organised as follows. In section~\ref{modified cosmology} we derive the modified Raychaudhuri and Friedmann equations and find their perturbative analytical solutions for some special cases. Section~\ref{numerical analysis} then contains numerical calculations of the perturbative corrections to the $\Lambda$CDM model. The domain of applicability of our approach is analysed in section~\ref{limitations}. Finally, section~\ref{discussion} sums up our results and discusses possible future directions.

\section{Modified Cosmology}
\label{modified cosmology}

In this section, we specialise our general quantum phenomenological gravitational equations to a cosmological model. The modifications we find will be relevant at scales near the Planck density but still large enough to consider just low-energy quantum gravity effects and a continuous spacetime. Of course, at late times, we expect to recover the classical cosmology up to negligible corrections.

First, we briefly summarize the origin of the quantum phenomenological equations of motion developed in the literature that we will use for the derivation of the cosmological solutions in this work. Then we obtain a modified Raychaudhuri equation for a universe filled with a (in general multi-component) perfect fluid. Finally, we develop a perturbative approach to solve that equation, finding a modified Friedman equation. Lastly, we obtain analytical solutions for some particular cases.

\subsection{Modified equations of gravitational dynamics}

The modified equations for gravitational dynamics we consider are derived from an entanglement equilibrium condition imposed on local causal diamonds constructed in every point of the spacetime. It has been shown that such condition encodes the  gravitational dynamics~\cite{Jacobson:2016}. The derivation is based on the maximal vacuum entanglement hypothesis, which establishes that at first order variation of matter fields and geometry from a vacuum maximally symmetric spacetime, the total entanglement entropy variation vanishes in a small geodesic ball of fixed volume~\cite{Jacobson:2016}. The variation of the total entropy is the sum of the variations of the matter entanglement entropy, $\delta S_{\text{m}}$, and entanglement entropy of the horizon corresponding to (modified) Bekenstein entropy~\cite{Jacobson:2016,Alonso:2020}, $\delta S_{\text{B}}$, i.e
\begin{equation}
\label{entropy ballance}
\delta S_{\text{B}} + \delta S_{\text{m}} =0.
\end{equation}

The modified entanglement entropy of the horizon is given by equation~\eqref{modi-entrop}. The matter entanglement entropy crossing the horizon for small perturbations of the vacuum can be expressed in terms of a local modular Hamiltonian proportional to the stress-energy tensor (see~\cite{Jacobson:2016} for details). Plugging expressions for both entropies to the equation~\eqref{entropy ballance} yields, after some nontrivial manipulations, the traceless modified equations governing gravitational dynamics
\begin{equation}
S_{\mu \nu} - D l_{\text{P}}^2 S_{\mu \lambda} S^{\lambda}_{\ \nu} + \frac{D l_P^2}{4} \left( R_{\kappa \lambda} R^{\kappa \lambda} - \frac{1}{4} R^2   \right) g_{\mu \nu} = \frac{8 \pi G}{c^4} \left( T_{\mu \nu} - \frac{1}{4} T g_{\mu \nu }\right), \label{eq::Mod_Eq_Mov}
\end{equation}
where $S_{\mu \nu}=R_{\mu \nu}-Rg_{\mu \nu}/4$, and $R_{\mu \nu}$ and $T_{\mu \nu}$ represent, respectively the Ricci and stress-energy tensors. For the full derivation of the modified equations, we refer the reader to the original paper~\cite{Alonso:2020}.

\subsection{Modified Raychaudhuri equation}
	
In this subsection, we obtain the modified Raychaudhuri equation governing the dynamics of the quantum phenomenological cosmological model. The modified equations introduce a correction term quadratic in the curvature tensors and proportional to parameter $D$ which quantifies the quantum gravitational modifications. It is easy to see that in the limit $D \rightarrow 0$ where the quantum effects vanish, we recover the standard traceless part of the Einstein equations. The tracelessness, together with some other features of these equations actually makes them a natural generalisation of the equations of motion of unimodular gravity, rather than general relativity. However, this distinction does not play a major role in the present work (for details, see~\cite{Alonso:2020}).

We then apply these general equations to the Friedmann-Lemaître-Robertson-Walker metric\footnote{Given the unimodular nature of the modified equations~\eqref{eq::Mod_Eq_Mov}, we should more appropriately use a unimodular form of the metric~\cite{Alonso:2020}. However, it can be checked that the Raychaudhuri equation derived from the standard and unimodular metric have exactly the same form (essentially due to conformal flatness of the metric). Therefore, we work with the standard version of the Friedmann-Lemaître-Robertson-Walker metric, as it is more familiar and somewhat easier to handle.}
\begin{eqnarray}
ds^2 =-c^2dt^2 + a^2(t) \left( \frac{dr^2}{1-kr^2} + r^2 \left(d\theta^2 +\sin^2{\theta} d\phi^2\right) \right) \label{eq::FLRW_metric},
\end{eqnarray}
where $a(t)$ is the scale factor and $k$ the curvature of the space.
	
To describe the matter content of the universe, we use the stress-energy tensor of a perfect fluid with two components: dust (non-relativistic matter) with the equation of state $p_{\text{m}}=0$ and radiation (relativistic matter) with $p_{\text{r}}=\rho_{\text{r}}c^2/3$. The stress-energy tensor reads
\begin{equation}
T_{\mu \nu} =\left(\rho + \frac{p}{c^2} \right) u_\mu u_\nu + \frac{p}{c^2}g_{\mu \nu},
\end{equation}
where the the energy and pressure density are the sums of the different perfect fluid components.
	
Since the equations of motion are traceless, they only yield one nontrivial condition. After some straightforward computations, we obtain for it
\begin{eqnarray}
\label{Raychaudhuri}
\Dot{H}-\frac{k}{a^2} - \frac{Dl_{\text{P}}^2}{c^2} \left(\Dot{H} - \frac{k}{a^2} \right)^2 = -4\pi G \left(\rho +\frac{p}{c^2}\right), \label{eq::Mod_Ray_with_curv}
\end{eqnarray}
where $H=\Dot{a}/a$ is the Hubble parameter. We recognize this expression as the Raychaudhuri equation with an extra term proportional to the correction parameter $D$. Taking the limit $D\to0$, we reproduce the classical Raychaudhuri equation, confirming the validity of this modified equation in the classical limit.
	
This modified Raychaudhuri equation is, in fact, an ordinary differential equation non-linear in the second derivatives of the scale factor. To obtain a solution, we develop in the following subsections a perturbative treatment in the non-linear term, appropriate when the gravitational fields are not too strong (i.e., everywhere except for the very beginning of the universe). This means assuming that the Hubble parameter corresponds to its classical value $H_{(0)}$ up to $O(l_{\text{P}}^2)$ terms and neglecting the $O(l_{\text{P}}^4)$ contributions. We discuss the range of validity of this approach in detail in section~\ref{limitations}.

\subsection{Modified Friedman equation: Perturbative solution}
	
In this subsection, we introduce a perturbative approach to solve the modified Raychaudhuri equation. For this purpose, we treat the term quadratic in $\dot{H}$ in the modified Raychaudhuri equation (\ref{eq::Mod_Ray_with_curv}) as a small correction. Then, we can perturbatively expand $H$ around its classical value $H_{(0)}$ as
\begin{equation}\label{eq::H_expansion}
H=H_{(0)}+H_{(1)}\frac{H_{(0)}^2l_{\text{P}}^2}{c^2}+O\left(\frac{H_{(0)}^4l_{\text{P}}^4}{c^4}\right), 
\end{equation}
and neglect the $O\left(l_{\text{P}}^4\right)$ terms. However, we keep the lowest nontrivial quantum modification of the order $H_{(1)} l_{\text{P}}^2$. This approximation corresponds to the regime in which $H_{(0)}^2$ (which can be thought of as the spacetime curvature scale) is significantly smaller than $c^2/l_{\text{P}}^2$, but still large enough that the term $H_{(0)}^2l_{\text{P}}^2/c^2$ is nonnegligible. Since the derivation of the modified gravitational equations~\eqref{eq::Mod_Eq_Mov} also neglects $O\left(l_{\text{P}}^4\right)$ terms, this is in fact just a consistency condition for our approach.
	
We can now approximate the correction term in the modified Raychaudhuri equation as
\begin{equation}
\frac{Dl_{\text{P}}^2}{c^2}\dot{H}^2=\frac{Dl_{\text{P}}^2}{c^2}\dot{H}_{(0)}^2+O\left(l_{\text{P}}^4\right).
\end{equation}
This yields the perturbative Raychaudhuri equation of the form
\begin{eqnarray}
\Dot{H} = -4\pi G \left(\rho +pc^2\right) + \frac{16\pi^2 G^2Dl_{\text{P}}^2}{c^2} \left(\rho +pc^2\right)^2 + \frac{k}{a^2}.
\end{eqnarray}
We consider a universe filled with a perfect fluid with two component, dust and radiation, because of its interest in cosmology. Using the equations of state, $p_i/c^2=\omega_i \rho_i$ with $\omega_{\text{r}} =1/3$ for the relativistic matter and $\omega_{\text{m}}=0$ for non-relativistic matter, we have for the perturbative Raychaudhuri equation 	
\begin{eqnarray}
\notag
\quad \Dot{H} &=& -4\pi G \sum_i \rho_i \left(\omega_i +1\right)  + \left(4\pi G\right)^2\frac{Dl_{\text{P}}^2}{c^2}   \left( \sum_i (\omega_i +1) \rho_i \right)^2 + \frac{k}{a^2}, \\
\Dot{H} &=& -4\pi G \sum_i \rho_i \left(\omega_i +1\right)  + \left(4\pi G\right)^2\frac{Dl_{\text{P}}^2}{c^2}   \left(\sum_i \left(\omega_i+1\right)^2 \rho_i^2 + 2 \sum_{i>j} \left(\omega_i \omega_j + \omega_i + \omega_j +1\right) \rho_i \rho_j \right) + \frac{k}{a^2}\label{eq::perturbative-Raychaudhuri equation}.
\end{eqnarray}
The modified equations of motion only yield one nontrivial condition for the metric, which does not suffice to fully determine it. This is a situation familiar from classical unimodular gravity, where one needs to add an equation for the divergence of the stress-energy tensor to obtain a complete system of equations~\cite{Alvarez:2021}. In principle, this divergence can be proportional to gradient of some scalar, i.e., $\nabla_{\mu}T^{\mu\nu} =\nabla^{\nu}\mathcal{J}$~\cite{Alvarez:2013}. If $\mathcal{J}\ne0$, the energy is not locally conserved. While this possibility has been considered in cosmology~\cite{Josset:2017}, we assume, for simplicity, that the local energy conservation holds, i.e., $\nabla_{\mu}T^{\mu\nu} = 0$. Moreover, we assume that the components of the fluid do not interact, and each of them is conserved separately (this is a reasonable assumption except for very early universe). This implies for each component
\begin{eqnarray}
\dot{\rho}_i + 3H\left(\rho_i + p_i\right) =0. \label{eq::local_energy_conservation}
\end{eqnarray}
Multiplying equation~\eqref{eq::perturbative-Raychaudhuri equation} by $2H$ and then integrating using the energy conservation for each fluid, yields the perturbative Friedman equation
\begin{eqnarray}
H^2 = \sum_i \frac{8\pi G}{3} \rho_i  -\sum_i \frac{16\pi^2 G^2}{3} \frac{Dl_{\text{P}}^2}{c^2} \left(\omega_i +1\right) \rho^2_i -  \frac{32\pi^2 G^2}{3}\frac{Dl_{\text{P}}^2}{c^2} \sum_{i>j} \frac{\left(\omega_i \omega_j + \omega_i + \omega_j +1\right)}{\omega_i +\omega_j +2} \rho_i \rho_j - \frac{k}{a^2} +  \Tilde{\Lambda}. \label{eq::Mod_Friedman_fluid}
\end{eqnarray}
 For our case of the two component fluid, the perturbative Friedman equation yields
\begin{eqnarray}
H^2 = \frac{8\pi G}{3} \left(\rho_m + \rho_r  \right)  - \frac{k}{a^2} - \frac{16\pi^2G^2 Dl_{\text{P}}^2}{3c^2} \left(\frac{4}{3} \rho_r^2 + \rho_m^2 + \frac{7}{5} \rho_m \rho_r \right) + \Tilde{\Lambda} .
\end{eqnarray}
	
In the limit of $D\to0$ this equation again reduces to the classical Friedman equation. The arbitrary integration constant $\tilde{\Lambda}$ plays the role of the cosmological constant, namely $\tilde{\Lambda}=\Lambda c^2/3$. Since, in classical unimodular gravity the cosmological constant also appears as an integration constant, being consistent with the unimodular nature of the modified equations we mentioned previously. It is convenient to write the Friedman equation in terms of dimensionless variables using the notation $\Omega_i = \left(8\pi G/3H_0^2\right)\rho_{i,0}$ for the relativistic and non-relativistic matter,with $i$ being r, m,  $\Omega_\text{k} = -k/\left(a_0^2H_0^2\right)$ for the curvature density and $\Omega_\Lambda = \Lambda c^2/\left(3H_0^2\right)$, where $a_0$, $H_0$, and $\rho_{i,0}$ denote the reference values of $a$, $H$, and $\rho_i$, typically taken to be the ones in the present day universe. The perturbative Friedmann equation in terms of these dimensionless variables reads
\begin{eqnarray}
\frac{H^2}{H_0^2} =  \left(\Omega_{\text{r}} \left(\frac{a_0}{a}\right)^4 + \Omega_{\text{m}} \left(\frac{a_0}{a} \right)^3 + \Omega_{\text{k}} \left( \frac{a_0}{a}\right)^2  + \Omega_\Lambda \right) - \frac{Dl_{\text{P}}^2}{12 c^2} H_0^2 \left(\frac{4}{3} \Omega_{\text{r}}^2  \left(\frac{a_0}{a} \right)^8 + \Omega_{\text{m}}^2  \left(\frac{a_0}{a} \right)^6  + \frac{7}{5}  \Omega_{\text{r}} \Omega_{\text{m}}  \left(\frac{a_0}{a} \right)^7  \right). \label{eq::Mod_Fried_matter_relative_density}
\end{eqnarray}
Therefore we can obtain a modified equation for the universe's content nowadays when we take into account phenomenological quantum-gravity modifications
\begin{eqnarray}
\Omega_{\text{r}} + \Omega_{\text{m}}  + \Omega_{\text{k}}  + \Omega_{\Lambda}  - \frac{Dl_{\text{P}}^2}{12 c^2} H_0^2 \left(\frac{4}{3} \Omega_{\text{r}}^2  + \Omega_{\text{m}}^2  + \frac{7}{5}  \Omega_{\text{r}} \Omega_{\text{m}}\right) =1.
\end{eqnarray}
Note that, in the present day universe, this equation differs from the established classical results for cosmology only negligibly, since the value of $Dl_{\text{P}}^2/c^2$ is very small, and the variation provided by the modified model can be included in the uncertainties of the measurements. The present day estimates for the dimensionless factors are $\Omega_{\text{r}}= 2.47\times 10^{-5} h^{-2}$ with $h=0.704 \pm 0.025$, $\Omega_{\text{m}}= 0.3111 \pm 0.0056 $, $\Omega_{\Lambda}  = 0.6889 \pm 0.0056$  and $\Omega_{\text{k}} = 0.0007 \pm 0.0019$ and for the Hubble parameter $H_0=67.66\pm 0.42 \text{ km } \text{s}^{-1} \text{Mpc}^{-1}$  \cite{Planck:2018vyg,Workman:2022ynf}.

\begin{figure}
    \centering
    \includegraphics[width=18cm]{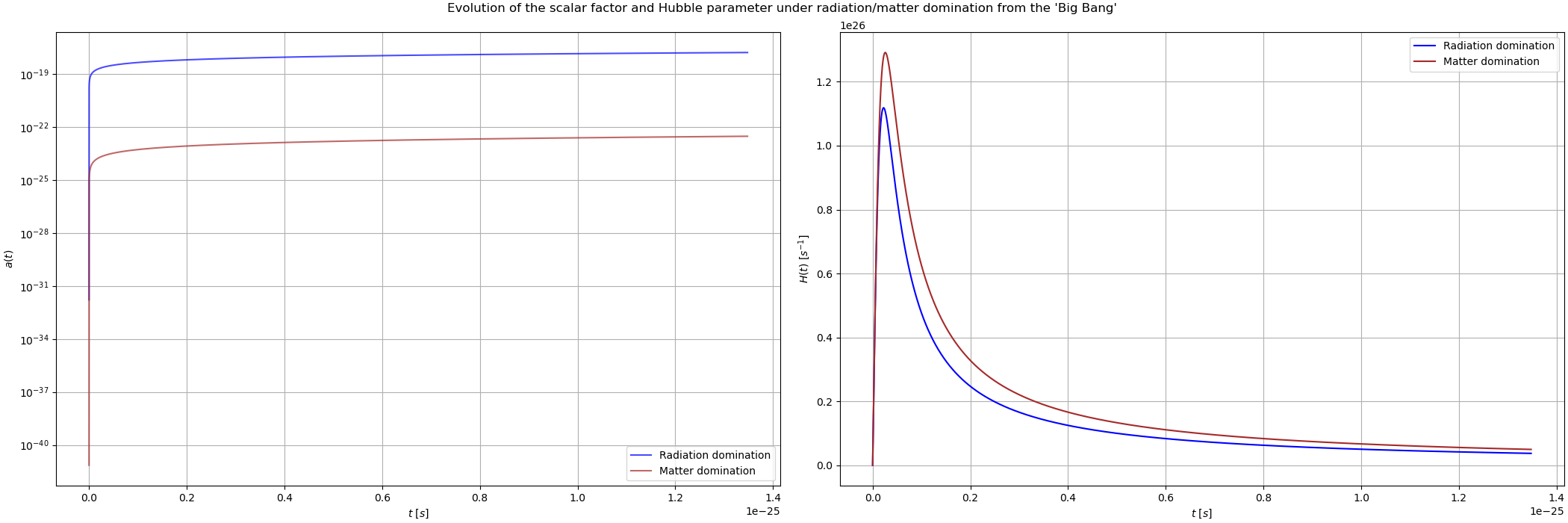} 
    
    \caption{On the left: Evolution of the scale factor under matter/radiation domination corresponding to the analytical solutions given in equations~\eqref{eq::scale_factor_m} and~\eqref{eq::scale_factor_r}. On the right: Evolution of the Hubble parameter under the same matter/radiation domination expressions~\eqref{matter_analytical}. For the figure, we fix $D=1$ and $a_0=1$.}
    \label{fig::Plot_Analytical}
\end{figure}

The perturbative Friedman equation~\eqref{eq::Mod_Fried_matter_relative_density} can be solved analytically, provided that we set $k=0$, $\Lambda=0$, and limit ourselves to a single component perfect fluid. In particular, we obtain analytical solutions for the evolution of the scale factor for dust domination and radiation domination. In the case of dust domination, i.e., neglecting radiation, the scale factor reads
\begin{eqnarray}
a_{\text{m}} (t) = a_0\left(\frac{9}{4} \Omega_{\text{m}} H_0^2 t^2 + \frac{Dl_{\text{P}}^2}{12c^2} \Omega_{\text{m}} H_0^2\right)^{1/3}.\label{eq::scale_factor_m}
\end{eqnarray}
It is direct to check that in the classical limit ($D\to0 $), the standard dependence $a\propto t^{2/3}$ for the dust domination period is recovered. For the case of radiation domination equation~\eqref{eq::Mod_Fried_matter_relative_density} yields
\begin{eqnarray}
a_{\text{r}} (t) =a_0 \left(4 \Omega_{\text{r}} H_0^2 t^2 + \frac{Dl_{\text{P}}^2}{9c^2} \Omega_{\text{r}} H_0^2 \right)^{1/4}.\label{eq::scale_factor_r}
\end{eqnarray}
In the classical limit, we again recover the standard solution for radiation domination period, $a \propto t^{1/2}$. In both cases, it is remarkable that for $t\to 0$ the initial singularity obtained in the standard cosmological model vanishes, and a minimum value for the scale factor appears. It depends on the correction term $Dl_{\text{P}}^2 /c^2$ and the actual values of $H_0$ and $\Omega_{\text{i}}$. It is also observable in a very short period where the scalar factor evolves very fast as it is shown in FIG.\ref{fig::Plot_Analytical}. This period could be interpreted as a kind of inflationary period. For the Hubble parameter, we have
\begin{eqnarray}
    H_{\text{m}}= \frac{2}{3} \frac{ t}{t^2 + \frac{D l_{\text{P}}^2}{27c^2}} \label{matter_analytical} \quad \text{and} \quad  H_{\text{r}}= \frac{t}{2t^2 + \frac{D l_{\text{P}}^2 }{18 c^2}}, \label{radiation_analytical}
\end{eqnarray}
In both cases, we recover the classical result for $D \to 0$, or for large $t$. We stress that around $t=0$, the Hubble parameter has a linear tendency with a value of zero at the origin. Nevertheless, we should take into account that we used a perturbative expansion  to obtain these results, and we should study the limitations of this model. The maximum of the Hubble parameter for radiation domination is obtained for $t_\text{max}= \sqrt{Dl_\text{P}^2/36c^2}$, this corresponds with $H_\text{max}=\sqrt{9c^2/4Dl_{\text{P}}^2}$ . We will discuss the limitations of this model in more detail in section~\ref{limitations}.

\section{Cyclic cosmology and numerical analysis}
\label{numerical analysis}

Once we have derived the modified equations and studying their perturbative solutions for dust and radiation analytically, we proceed to carry out a numerical analysis that allows us to address more complicated and realistic settings that cannot be studied analytically.

First of all, we need to confirm that our model reproduces the standard model of cosmology at late times. For this purpose, we use a numerical analysis based on the Forth Order Runge-Kutta Method to solve the modified Raychaudhuri equation \eqref{eq::Mod_Ray_with_curv} taking into account all of the parameters. This method permits us to study the evolution of the system forwards and backward in time from today and to use different sets of parameters as the initial conditions. The parameters chosen for the system are the density parameters ($\Omega_{\text{r}},\Omega_{\text{m}},\Omega_{\Lambda},\Omega_{\text{k}}$), the today's value for the scale factor (fixed to $a_0=1$ by convenience) and the dimensionless parameter $D$. The theoretical predictions for this parameter are suggested to be in general about order of one~\cite{Alonso:2020}, thus we fix $D=1$ just for showing graphically the results. We also choose an adjusted time scale to study the system. For convenience, we use the Hubble scale for time, fixing thus today's Hubble parameter $H_0=1$. Moreover, we set the beginning of time parameter so that $H_0$ corresponds to $\tau=0$.  We did a change of variable to respect the cosmic time $t$, $\tau= t + t_0$, where $t_0$ is the age of the universe.

\begin{figure}
    \centering
    \includegraphics[width=18cm]{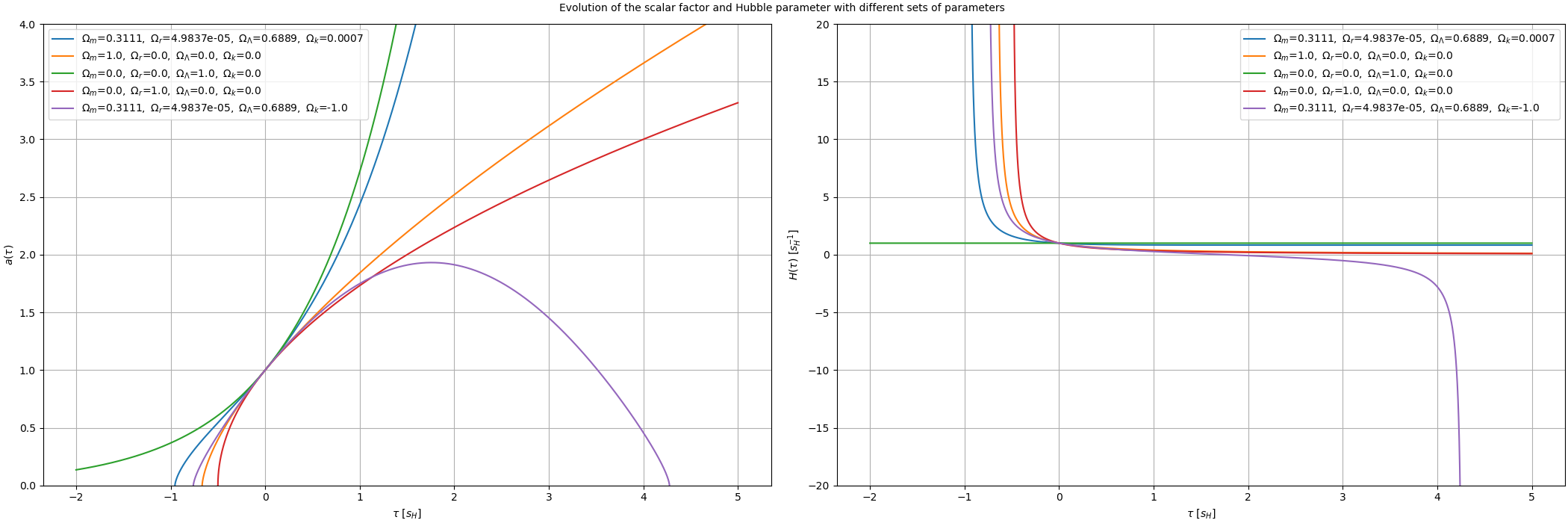} 
    
    \caption{ On the left: Evolution of the scale factor under different sets of parameters in Hubble scale. On the right: Evolution of the Hubble parameter under the same sets of parameters. For the figure, we fix $D=1$ and $a_0=1$. We set the time coordinate so that the (present day) values $a_0$, $H_0$ correspond to $\tau=0$}
    \label{fig::plot_numerical_resolution}
\end{figure}

In order to depict FIG. \ref{fig::plot_numerical_resolution}  we used different sets of parameters: today's parameters based on the $\Lambda$CMB model, Einstein-de Sitter configuration (only matter component), de Sitter configuration, only radiation component and finally a closed universe ($k=1$) with the today's parameters based on the $\Lambda$CMB model too. We stress that the impact of the modification of the Friedman equation is almost negligible on large scales, as the Hubble scale, since the evolution is quite similar to the evolution without modifications ($D\to 0$). In the figure we can easily see a change of behaviour in the expansion of scale factor and Hubble parameter near the origin.

In the last section, we pointed out a minimum positive value for the scale factor using a perturbative treatment for the modified Raychaudhuri equation \eqref{eq::Mod_Ray_with_curv}. The emergence of this minimal value and the avoidance of the singularity are consequences of reaching a zero value for the Hubble parameter at the moment of the "Big Bang". This point can represent a bounce between a contracting phase of the universe ($H<0$) to the expanding one ($H>0$).The emergence of a cyclic universe appears immediately if the universe is not flat, concretely in a closed universe with positive curvature ($k>0$ and $\Omega_{\text{k}} <0$) where a turning point arrives as in the standard cosmology to reach the cyclical evolution. The "Big Bang" singularity is replaced by a regular bounce between the expansion and contraction phases, thus giving rise to cosmic bounce. This last idea will be developed in detail in the next section.

\section{Domain of validity of the model}
\label{limitations}

In this section we want to bring the attention to the range of validity of the phenomenological model we have developed throughout the paper. As we have previously pointed out, the regular bounce we observed for $D>0$ is a perturbative result obtained when neglecting all $O\left(l_{\text{P}}^2\right)$ terms. This is consistent with the original derivation of the modified equations for gravitational dynamics resulting in the expression~\eqref{eq::Mod_Eq_Mov} which also treats the $O\left(l_{\text{P}}^4\right)$ contributions to entropy as negligible. Nevertheless, we may ask whether the modified Raychaudhuri equation we derived still has validity even when we formally treat it as a nonperturbative, exact equation. In that case, we obtain for $\dot{H}$ (the sign of the square root is chosen so that we reproduce the classical Raychaudhuri equation in the limit $D\to0$)
\begin{equation}
\label{nonperturbative}
\frac{2Dl_{\text{P}}^2}{c^2}\left(\dot{H}-\frac{k}{a^2}\right) = 1-\sqrt{1+\frac{16\pi GDl_{\text{P}}^2}{c^2}\left(\rho+\frac{p}{c^2}\right)}. 
\end{equation}
One can see in this expression, on the one side, that for $D>0$ we have $\dot{H}<0$ at all times and a bounce is therefore not possible. On the other side, for $D<0$ we again have $\dot{H}<0$, but the square root develops a zero when the matter density reaches the Planck scale. At this point, $\dot{H}$ attains an imaginary part, making the evolution of $H$ ill-defined. Since $\dot{H}<0$ when this point is reached, it cannot correspond to a bounce. This behaviour strongly suggests that our equations can be trusted only perturbatively, and all the $O\left(l_{\text{P}}^4\right)$ corrections should be consistently neglected.

Nevertheless, we can use equation~\eqref{nonperturbative} to estimate the limit of applicability of the perturbative approach. Expanding it up to the $O\left(l_{\text{P}}^4\right)$ order yields
\begin{equation}
\dot{H}-\frac{k}{a^2} = -4\pi G\left(\rho+\frac{p}{c^2}\right)+\frac{16\pi^2G^2Dl_{\text{P}}^2}{c^2}\left(\rho+\frac{p}{c^2}\right)^2-\frac{128\pi^3G^3D^2l_{\text{P}}^4}{c^4}\left(\rho+\frac{p}{c^2}\right)^3+O\left(l_{\text{P}}^6\right). 
\end{equation}
We see here that the $O\left(l_{\text{P}}^4\right)$ contribution has an opposite sign as the $O\left(l_{\text{P}}^2\right)$. The absolute values of both contributions become equal at
\begin{equation}
\rho+\frac{p}{c^2}=\frac{c^2}{8\pi GDl_{\text{P}}^2}=\frac{\rho_{\text{P}}}{8\pi D},
\end{equation}
where $\rho_{\text{P}}=c^2/\left(Gl_{\text{P}}^2\right)$ denotes the Planck density. At this density, a perturbative expansion neglecting $O\left(l_{\text{P}}^4\right)$ certainly fails, as these become equally important as the $O\left(l_{\text{P}}^2\right)$ ones. Since the bounce corresponds to
\begin{equation}
\rho+\frac{p}{c^2}=\frac{\rho_{\text{P}}}{4\pi D}>\frac{\rho_{\text{P}}}{8\pi D},
\end{equation}
it already lies beyond the limit of applicability of our perturbative expansion. Therefore, the presence of the bounce cannot be confirmed by our model, although, for $D>0$, the modified dynamics do suggest it. To provide stronger results, the $O\left(l_{\text{P}}^4\right)$ (or higher) terms would have to be included already in deriving the general modified equations for gravity~\eqref{eq::Mod_Eq_Mov}. This would in turn introduce additional undetermined parameters besides $D$, making the predictions we obtain dependent on their precise interplay. Then, the predictive power of our model would not be improved by such approach. Instead, it appears more worthwhile to apply the modified equations to phenomena that lie still in their regime of validity, but can be already affected by the Planck scale corrections we introduced.

\section{Discussion}
\label{discussion}

\begin{figure}
    \centering
    \includegraphics[width=18cm]{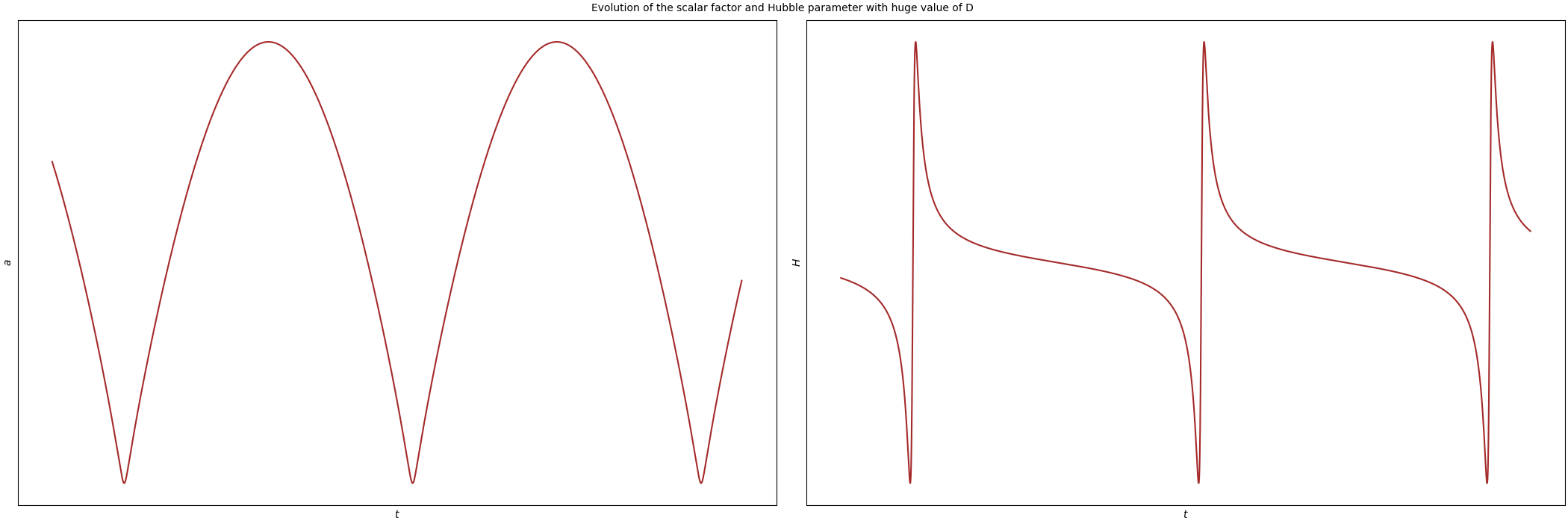} 
    
    \caption{ On the left: Qualitative evolution of the scale factor under the hypothesis of cyclic universe. On the right: Qualitative evolution of the Hubble parameter under the hypothesis of cyclic universe}
    \label{fig::cyclic_universe_plot}
\end{figure}

Starting from a general phenomenological approach based on thermodynamics of spacetime, we have obtained low energy quantum gravitational modifications to the homogeneous, isotropic universe filled with a perfect fluid. For the case of single component perfect fluids (dust or radiation) we found perturbative analytical solutions. However, the case including dust and radiation together needed to be studied numerically. Both analytical and numerical results show that the late time evolution corresponds almost precisely to the classical cosmology (this is the natural, since the modifications we introduced are proportional to Planck length squared). In the early universe, the modifications we introduced suggest the replacement of the Big Bang singularity by a regular bounce provided that $D>0$. If the universe starts contracting again at late times (e.g. due to having a positive spatial curvature), this implies a cyclic cosmological model, in FIG.\ref{fig::cyclic_universe_plot} we can see the qualitative evolution expected for this cyclic universe. These results are also consistent with the predictions of loop quantum cosmology. This is expected, since our perturbative  modified Raychaudhuri equation is very similar to the one valid for the effective dynamics of loop quantum cosmology.

While the dynamics of our model suggests the resolution of the Big Bang singularity, this result is obtained perturbatively. Therefore, we cannot conclusively show that the singularity resolution indeed occurs without including higher order terms in the Planck length already on the level of the derivation of the general modified equations (we stress that this is a general issue affecting most of the quantum cosmological approaches, not particular to our model). Nevertheless, the present form of the modified cosmological equations can yield novel and definite results in the regime at which gravity is very strong, but still sufficiently below Planck scale. 

The modifications to the cosmological evolution we derived from a local thermodynamic approach are also qualitatively consistent with other thermodynamic approaches in the literature which instead consider thermodynamics of the global apparent horizon~\cite{Salah_2017,Hern_ndez_Almada_2022}. These approaches introduce corrections to Bekenstein entropy from modified GUP~\cite{Salah_2017} or use different definitions for the entropy~\cite{Hern_ndez_Almada_2022}. In both cases, they found the emergence of a cosmic bounce and study the existence of the resulting cyclic universe.

In particular for future projects, our model can have implications for the inflation period, for the cosmological perturbations and for cosmic microwave background. Moreover, it is of interest to apply it to anisotropic cosmological models. All of these questions were studied in the context of loop quantum cosmology~\cite{Martin-Benito:2021,deBlas:2016,Han:2020,Ashtekar:2021,Wilson-Ewing:2010,Martin-Benito:2010} and in a future work we will study them in our model checking as well the equivalence with the particular previous results.

\section*{Acknowledgments}

AA-S is supported by the ERC Advanced Grant No. 740209. ML is supported by the Charles University Grant Agency project No. GAUK 297721. AV-B acknowledges the hospitality of Albert Einstein Institute during the development of the work giving rise to this project. This work is also partially supported by the Spanish Government through Project. No. MICINN PID2020-118159GB-C44.

\bibliography{apssamp}

\end{document}